\documentclass[12pt,paper]{article}
\usepackage{amsmath}
\usepackage{verbatim} 
\usepackage{amsfonts}
\usepackage{amssymb}
\usepackage{amsthm}
\usepackage{graphics}
\usepackage{graphicx}
\usepackage{theorem}
\usepackage{bm}
\usepackage{apacite}
\usepackage[authoryear]{natbib}
\usepackage[english]{babel}
\usepackage{caption}
\usepackage{subcaption}

\begin{document}
\selectlanguage{english}
\title{A fast Mixed Model B-splines algorithm }
\author{Martin P. Boer 
        \\ Biometris WUR \\ Wageningen \\
       The Netherlands \vspace{2mm} \\
       {\tt martin.boer@wur.nl}}
\date{\today}

\maketitle

\section*{Abstract}
A fast algorithm for B-splines in mixed models is presented. B-splines have local support and are computational attractive, because the corresponding matrices are sparse. A key element of the new algorithm is that the local character of B-splines is preserved, while in other existing methods this local character is lost. The computation time for the fast algorithm is linear in the number of B-splines, while computation time scales cubically for existing transformations. 
\\

\newpage
\section{Introduction}\label{sec:intro}
Penalized regression using B-splines can be computationally efficient, because of the local character of B-splines. The corresponding linear equations are sparse and can be solved quickly. However, the main problem is to find the optimal value for the penalty parameter. A good way to approach this problem is to use mixed models and restricted maximum likelihood \citep[REML;][]{Patterson1971}. Several methods have been proposed to transform the original penalized B-spline model to a mixed model \citep{Currie2002,Lee2011}. A problem with existing transformations to mixed models is that the local character of the B-splines is lost, which reduces the computational efficiency. For relatively small datasets this is not a major issue. However, for long time series, for example with measurements every five minutes for several months, the computational efficiency becomes quite important. 

In this paper I present a new transformation to a mixed model. This model is closely related to the transformation proposed by \cite{Currie2002}. However, the computation time in the transformation of \cite{Currie2002} increases cubically in  the number of B-splines, while for the new transformation the computation time increases linearly in time, using sparse matrix algebra \citep{Furrer2010}. One of the key elements of the proposed algorithm is that the transformation preserves the local character of B-splines, and all the equations can be solved quickly.
     
The paper is organized as follows. In Section~\ref{sec:Bsplines} relevant information about B-splines is given. In Section~(\ref{sec:PsplinesMixedModels}) first the P-spline model \citep{Eilers1996} is described and the transformation to mixed models by \cite{Currie2002} is stated. The new transformation is presented, and details for a sparse mixed model formulation are given. In Section~\ref{sec:Rcode)} the R-code is briefly described and a comparison is made between the computation time of the new method and the transformation by \cite{Currie2002}.

\section{B-splines}\label{sec:Bsplines}
In this preliminary section, a few relevant details about B-splines are given. For a detailed overview of B-splines, see for example \cite{Boor1978} and \cite{Hastie2009}. B-splines have local support. This property is important and can speedup calculations considerably. To illustrate the idea of the local support, see Figure~1, with quadratic (i.e.\ degree $q=2$) B-splines. Throughout the paper we will assume equal distance between the splines, denoted by $h$. In the example presented in Figure~1 the distance is unity, $h=1$. The number of B-splines will be denoted by $m$. The first quadratic B-spline, $B_{1,2}(x)$ is zero outside the interval $[-2,1]$. The last one, $B_{12,2}(x)$, is zero outside the interval $[9,12]$. So, for this example, there are $m=12$ quadratic B-splines which define the B-spline basis for the domain $[x_{\text{min}},x_{\text{max}}]=[0,10]$ of interest.  

\begin{figure}[t]
\begin{center}
\includegraphics[scale=0.9]{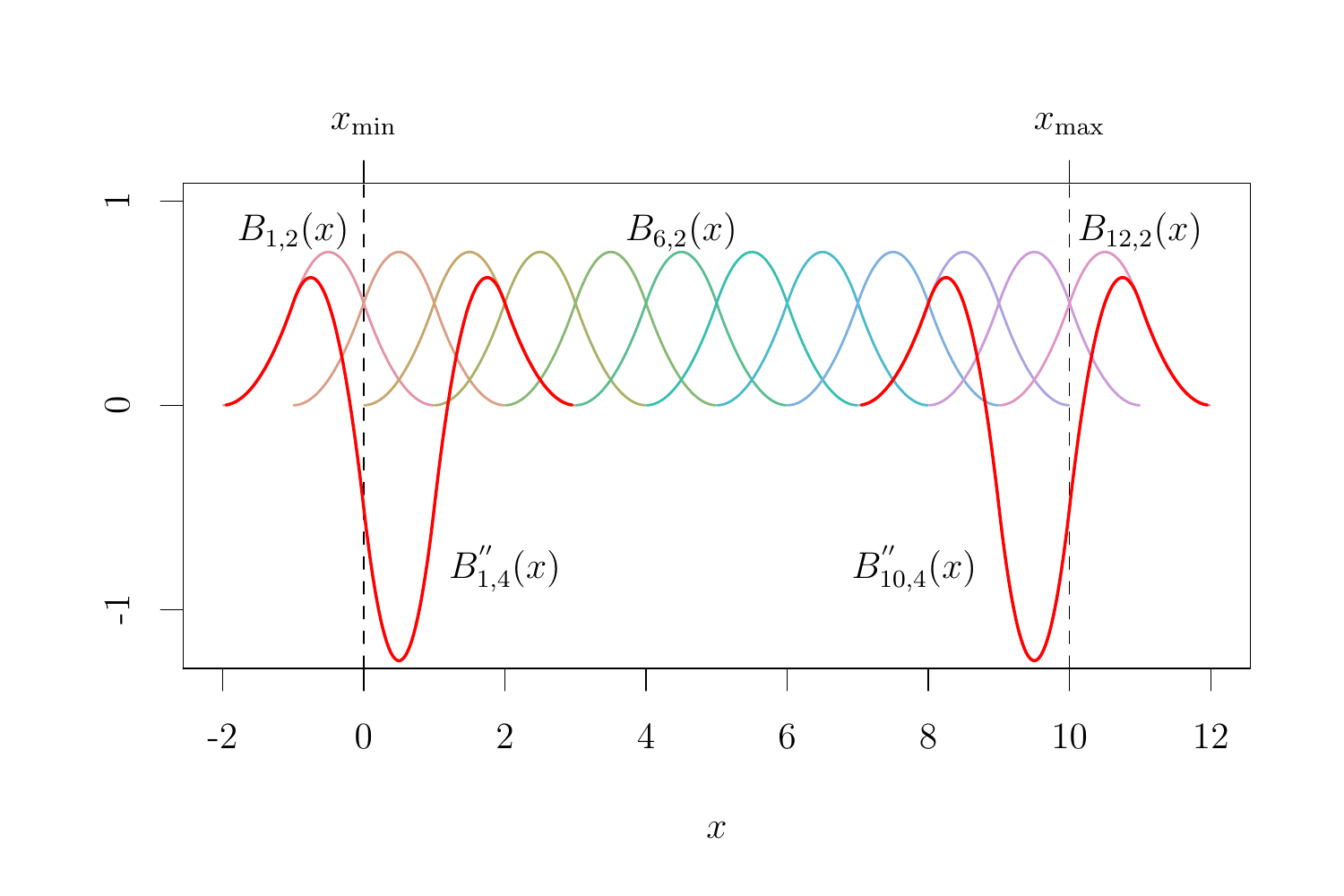}
\caption{{\small Quadratic B-spline basis for the interval $[x_{\text{min}},x_{\text{max}}]=[0,10]$, formed by $B_{1,2}(x)$, $B_{2,2}(x)$, ... , $B_{12,2}(x)$. The distance between the splines is unity. The red curves are second-order derivatives of quartic B-splines; for clarity only the first and last one are shown. The second-order derivatives of quartic B-splines can be constructed from quadratic B-splines, see equation~(\ref{eqn:Bsderiv}).
}}
\end{center}
\end{figure}

The second derivative of B-splines is given by \citep{Boor1978}:
\begin{equation}
  h^2 B^{''}_{j,q}(x) = B_{j,q-2}(x) - 2 B_{j+1,q-2}(x) + B_{j+2,q-2}(x) \;. \label{eqn:Bsderiv} 
\end{equation} 
The second derivative is illustrated in Figure~1. The red curves are second-order derivatives of quartic (fourth-degree) B-splines. As can be seen from equation~(\ref{eqn:Bsderiv}) it can represented as a linear combination of three quadratic B-splines. 

\section{P-splines and Mixed Models}\label{sec:PsplinesMixedModels}
In this section we will give a brief description of P-splines \citep{Eilers1996}. Let $n$ be the number of observations. Suppose the variable $\mathbf{y} = (y_1,\ldots,y_n)'$ depends smoothly on the variable $\mathbf{x} = (x_1,\ldots,x_n)'$. Let $\mathbf{B}  =(B_{1,2}(\mathbf{x}),\ldots,B_{m,2}(\mathbf{x}))$ be a $n \times m$ matrix, and $\mathbf{a} = (a_1,a_2,\ldots,a_m)'$ be a vector of regression coefficients. Then the following objective function to be minimized can be defined:
\begin{equation}\label{eqn:Psplines}
    S(\mathbf{a}) = 
    (\mathbf{y} - \mathbf{B} \mathbf{a})'(\mathbf{y} - \mathbf{B} \mathbf{a}) 
      + \lambda \; \mathbf{a}' \mathbf{D}' \mathbf{D} \mathbf{a}
   \;,
\end{equation}
where $\lambda>0$ is a penalty or regularization parameter, and $\mathbf{D}$ is an $(m-2) \times m$ second-order difference matrix \citep[see e.g.][]{Eilers1996}.

\cite{Currie2002} showed that equation~(\ref{eqn:Psplines}) can be reformulated as a mixed model: 
\begin{equation}\label{eqn:mixedmodel}
   \mathbf{y} = \mathbf{X b} + \mathbf{Z u} + \mathbf{e} \;, \quad
   \mathbf{u} \backsim N(\mathbf{0},\frac{1}{\lambda}\mathbf{Q}^{-1} \sigma^2) \;,
   \quad \mathbf{e} \backsim  N(\mathbf{0},\mathbf{I}\sigma^2) \;,
\end{equation}
where $\mathbf{X}$ and $\mathbf{Z}$ are design matrices, $\mathbf{Q}$ is a precision matrix, $\mathbf{b}=(b_0,b_1)'$ are the fixed effects, $\mathbf{u} = (u_1,u_2,\ldots,u_{m-2})'$ the random effects, $\mathbf{e}$ is the residual error, and $\sigma^2$ is the residual variance. 

\cite{Currie2002} used the following transformation:
\begin{equation}\label{eqn:currietransform}
   \mathbf{a} = \mathbf{G} \mathbf{b} + \mathbf{D}' (\mathbf{D} \mathbf{D}')^{-1} \mathbf{u} \; ,
\end{equation}
where $\mathbf{G}$ is an $m \times 2$ matrix with columns $\mathbf{g}_0 = (1,1,\ldots,1)'$ and $\mathbf{g}_1 = (1,2,\ldots,m)'$. This transformation gives the following expressions for the design matrices and the  precision matrix:
\begin{equation}\label{eqn:Z_Currie}
      \mathbf{X} = \mathbf{B} \mathbf{G}  \;, \quad
      \mathbf{Z} = \mathbf{B} \mathbf{D}' (\mathbf{D} \mathbf{D}')^{-1}  \;, \quad
      \mathbf{Q} = \mathbf{I} \;.
\end{equation}      

The mixed model equations \citep{Henderson1963} corresponding to equation~(\ref{eqn:mixedmodel}) are given by:
\begin{equation}\label{eqn:mme}
    \begin{pmatrix}
        \mathbf{X'X} & \mathbf{X'Z} \\
        \mathbf{Z'X} & \mathbf{Z'Z} + \lambda \mathbf{Q}
    \end{pmatrix}
    \begin{pmatrix}
    	\widehat{\mathbf{b}} \\ \widehat{\mathbf{u}} 
       \end{pmatrix} =
  \begin{pmatrix}
       \mathbf{X'y} \\ \mathbf{Z'y}
    \end{pmatrix} \;.
\end{equation}
The coefficient matrix $\mathbf{C}_{\lambda}$ in equation~(\ref{eqn:mme}) is given by:
\begin{equation}\label{eqn:mme_coefmatrix}
   \mathbf{C}_{\lambda} = 
    \begin{pmatrix}
        \mathbf{X'X} & \mathbf{X'Z} \\
        \mathbf{Z'X} & \mathbf{Z'Z} + \lambda \mathbf{Q} 
    \end{pmatrix} \;.
\end{equation}
This coefficient matrix is dense, since the local character of the B-splines has been destroyed by equation~(\ref{eqn:Z_Currie}). This implies that the computation complexity for solving equation~(\ref{eqn:mme}) is $\mathcal{O}(m^3)$.

The following transformation preserves the local character of the B-splines:
\begin{equation}\label{eqn:mmb_transform}
   \mathbf{a} = \mathbf{G} \mathbf{b} + \mathbf{D}' \mathbf{u} \;.
\end{equation}
Figure~1 illustrates the underlying idea of this transformation. The quadratic B-splines basis consists of $m=12$ B-splines. This quadratic B-spline basis is transformed to a second-order derivative quartic B-splines basis of $m-2=10$ B-splines, plus a parameter for intercept $b_0$ and linear trend $b_1$. The second-order derivative quartic B-splines can be constructed from quadratic B-splines by second-order differencing.
Using the new transformation the design and precision matrices are given by:
\begin{equation}\label{eqn:mmb_model}
      \mathbf{X} = \mathbf{B} \mathbf{G}  \;, \quad 
      \mathbf{Z} = \mathbf{B} \mathbf{D}' \;, \quad
      \mathbf{Q} = \mathbf{D} \mathbf{D}' \mathbf{D} \mathbf{D}' \;.
\end{equation}      
Let us refer to equations~(\ref{eqn:mixedmodel}) and~(\ref{eqn:mmb_model}) as a Mixed Model of B-splines (MMB), since it uses the B-splines directly as building blocks for the mixed model. The matrix $\mathbf{Z}'\mathbf{Z}+\lambda \mathbf{Q}$ has bandwidth $4$. This implies that $\mathbf{C}_\lambda$ is sparse and computation complexity has been reduced to $\mathcal{O}(m)$. An efficient way to calculate the REML profile log likelihood \citep{Gilmour1995, Crainiceanu2004,Searle2009} is given by the following four steps :

\begin{enumerate}
 \item Sparse Cholesky factorization \citep{Furrer2010}: $\mathbf{C}_\lambda = \mathbf{U}_{\lambda} \mathbf{U}^{'}_{\lambda}$, where $\mathbf{U}_{\lambda}$ is an upper-triagonal matrix.

 \item Forward-solve and back-solve \citep{Furrer2010}, with $\mathbf{w}$ a vector of length $m$:
           \begin{equation}
               \mathbf{U}_\lambda \mathbf{w} =   \begin{pmatrix}
       		\mathbf{X'y} \\ \mathbf{Z'y}
    		\end{pmatrix}  \;,
    		\quad
			\mathbf{U}^{'}_{\lambda}     
	    \begin{pmatrix}
    	\widehat{\mathbf{b}} \\ \widehat{\mathbf{u}} 
       \end{pmatrix} = \mathbf{w} \;.
          \end{equation}
 \item Calculate $\hat{\sigma}^2$ \citep{Johnson1995}, $p$ is the dimension of the fixed effects:
          \begin{equation}
    \hat{\sigma}^2 =  
    \frac{\mathbf{y}'\mathbf{y}
    -\widehat{\mathbf{b}}'\mathbf{X}' \mathbf{y}  
    -\widehat{\mathbf{u}}'\mathbf{Z}' \mathbf{y} 
    }{n-p} \;.
		\end{equation}
 \item REML log profile likelihood \citep{Gilmour1995, Crainiceanu2004,Searle2009}:
      \begin{equation}
  L(\lambda) = -\frac{1}{2} \left( 2 \log |\mathbf{U}_\lambda| 
          - (m-p) \log \lambda 
          + (n-p) \log \hat{\sigma}^2
           + C \right) \;,
     \end{equation} 
     where $C$ is a constant: $C = n-p - \log |\mathbf{Q}|$.
\end{enumerate}
A one-dimensional optimization algorithm can be used to find the maximum for $L(\lambda)$. The computation time is linear in $m$.  

\section{R-package MMBsplines}\label{sec:Rcode)}
An R-package, {\bf MMBsplines}, is available at GitHub: 

\url{https://github.com/martinboer/MMBsplines.git}. 

The sparse matrix calculations are done with the {\bf spam} package \cite{Furrer2010}. The B-splines are constructed with {\tt splineDesign()} of the {\bf splines} library. 

The following example code sets some parameter values and runs the simulations:
{\small
\begin{verbatim}
    nobs = 1000; xmin = 0; xmax = 10
    set.seed(949030)
    sim.fun = function(x) { return(3.0 + 0.1*x + sin(2*pi*x))}
    x = runif(nobs, min = xmin, max = xmax)
    y = sim.fun(x) + 0.5*rnorm(nobs)
\end{verbatim}}
A fit to the data on a small grid can be obtained as follows, using $m=100$ quadratic B-splines:
{\small
\begin{verbatim}
    obj = MMBsplines(x, y, xmin, xmax, nseg = 100)
    x0 = seq(xmin, xmax, by=0.01)
    yhat = predict(obj, x0)
    ylin = predict(obj, x0, linear = TRUE)
    ysim = sim.fun(x0)    
\end{verbatim}}

\begin{figure}[h]
\begin{center}
  \includegraphics[scale=0.6]{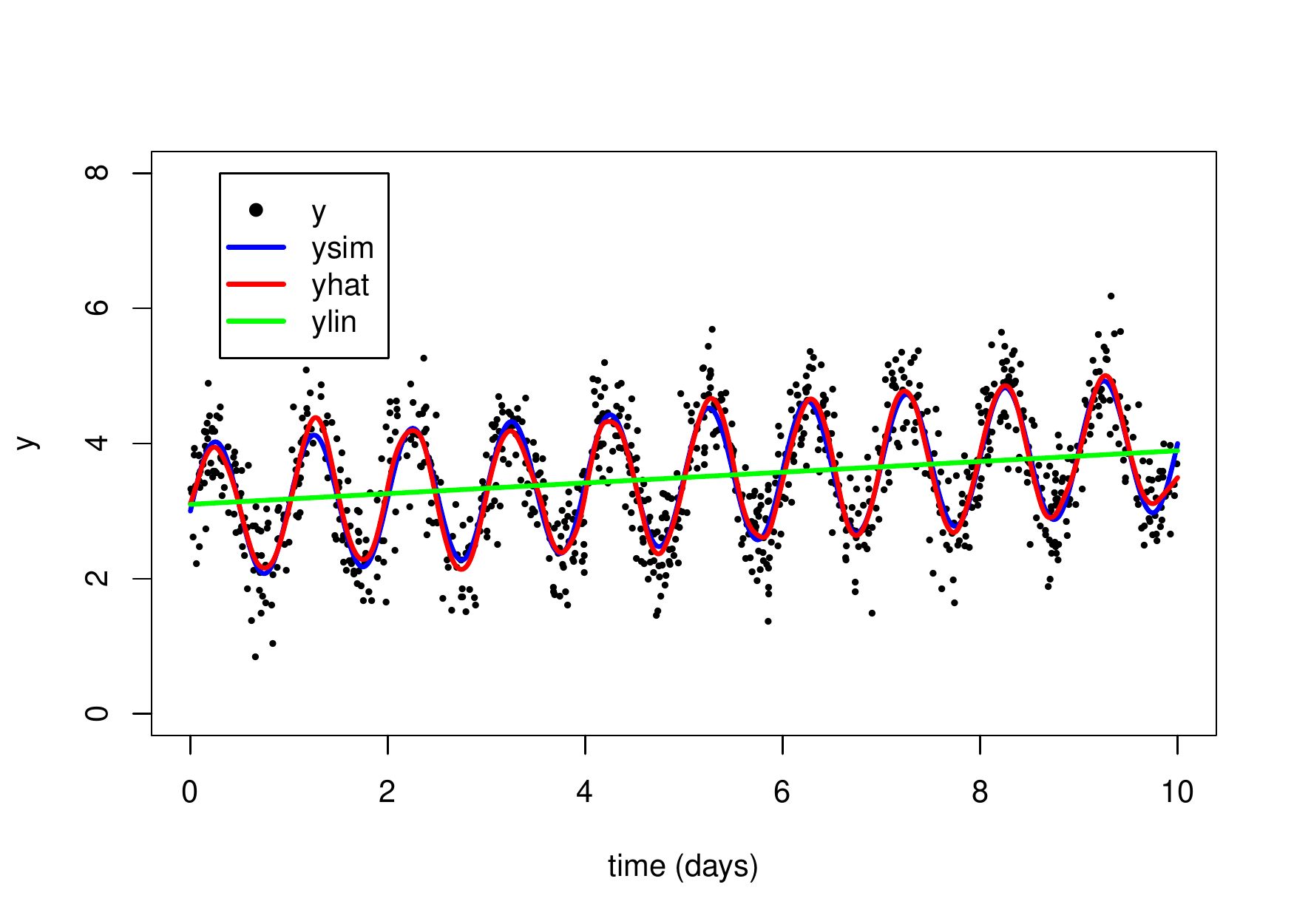}
\end{center}
\caption{{\small Fit of the simulated data using {\bf MMBsplines}, with $\lambda_{\text{max}}=1.33$. The blue line is the true simulated line, the red line is the fitted value. The green line is the linear trend.}}\label{fig:sim10days}
\end{figure}

Figure~\ref{fig:sim10days} shows the result, with $\lambda_{\text{max}}=1.33$. 

The Currie and Durban transformation can be run by setting the {\tt sparse } argument to {\tt FALSE}:
\begin{verbatim}
    obj = MMBsplines(x, y, xmin, xmax, nseg = 100, sparse = FALSE)
\end{verbatim}}
For $m=100$, as in Figure~\ref{fig:sim10days}, the differences in computation time are small. If we increase the length of the simulated time series, with a fixed stepsize $h=0.1$, the advantage of the MMB-splines method becomes clear, see Figure~3. As expected the Currie and Durban transformation computation time increases cubical in the number of B-splines, computatation time for MMB-splines is linear in $m$.

\begin{figure}[t]
\begin{center}
  \includegraphics[scale=0.6]{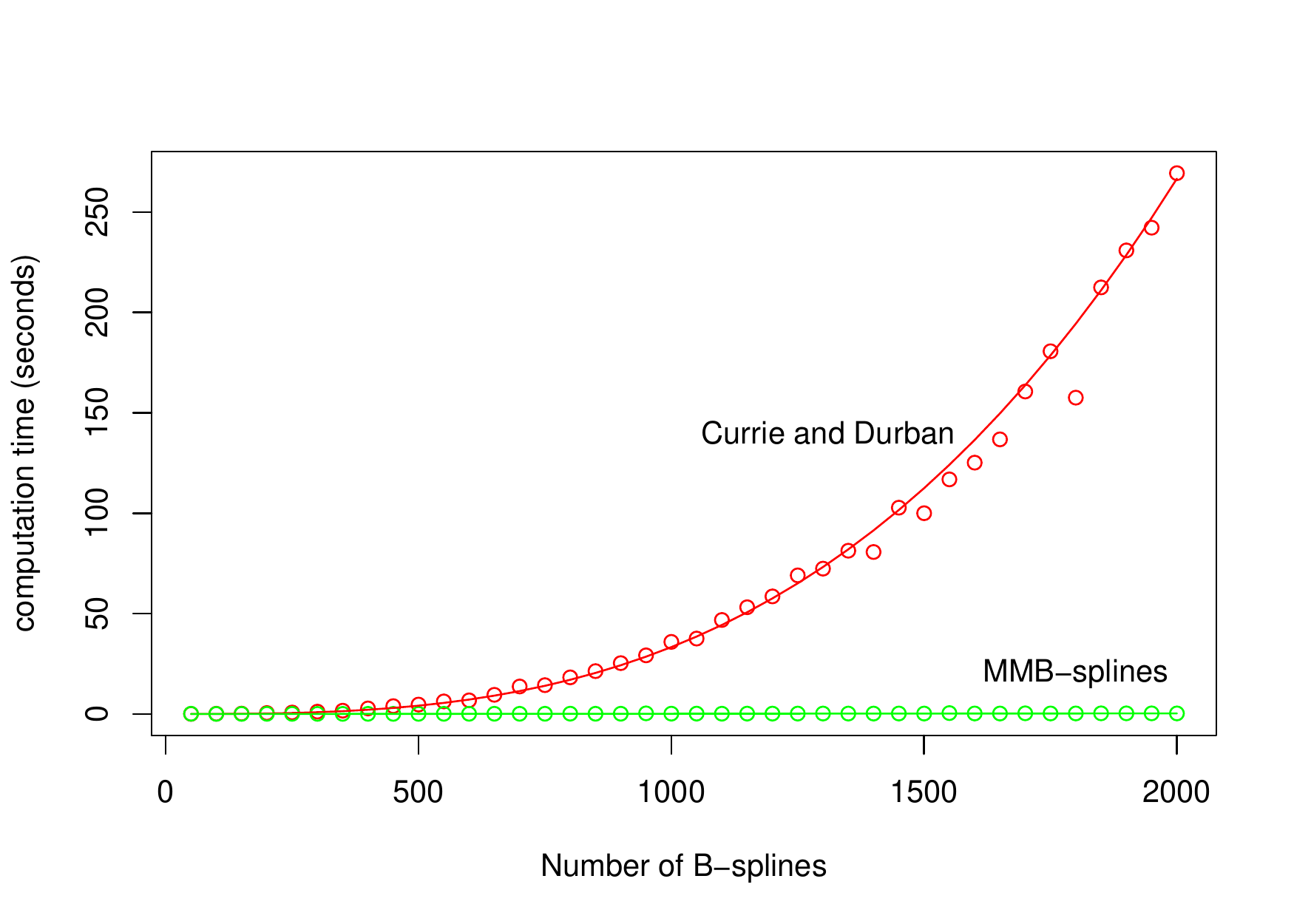}
\end{center}
\caption{{\small Comparison of computation times. The computation time for the Currie and Durban transformation is cubical in the number of B-splines. The computation time for MMB-splines is linear in the number of B-splines.}}
\end{figure}

\section{Conclusion}\label{sec:Discussion)}
The MMB-splines method presented in this paper seems to be an attractive way to use B-splines in mixed models. The method was only presented for quadratic splines, but also cubical or higher-degree B-splines could have been used. Other generalizations are also possible, for example extension to multiple penalties \citep{Currie2002} or multiple dimensions \citep{Rodriguez-Alvarez2014}. 

\section*{Acknowledgments}
I am indebted to Hugo van den Berg for useful comments on earlier drafts. I would also like to thank Paul Eilers, for explaining to me the local character of B-splines, and many valuable discussions. I would like to thank Cajo ter Braak and Willem Kruijer for valuable discussions and corrections of earlier versions of the paper.

\bibliographystyle{apacite}
\bibliography{mbnotes}

\end{document}